\documentclass[aps, prd ]{revtex4}

\usepackage{amsmath,amsfonts,amscd,amssymb,epsf, color}
\usepackage[small]{caption}

\newcommand{\pa}{\partial}

\newcommand{\vep}{\varepsilon}

\begin{document}

\title{Zero and finite temperature Casimir effect of massive vector field between  real metals}
\author{L. P. Teo}
 \email{LeePeng.Teo@nottingham.edu.my}
\affiliation{Department of Applied Mathematics, Faculty of Engineering, University of Nottingham Malaysia Campus, Jalan Broga, 43500, Semenyih, Selangor Darul Ehsan, Malaysia.}
\begin{abstract}
We consider the Casimir effect of a massive vector field between two semi-infinite dielectric slabs. We first derive the generalization of the Lifshitz formula that gives the Casimir interaction energy of two magnetodielectric slabs separated by a magnetodielectric medium due to the vacuum fluctuations of a   massive vector field. We then discuss the asymptotic behaviors of the Casimir energy and the Casimir force in various limits, such as low temperature, high temperature, small mass, large mass, up to the first order in the finite conductivity correction, for two real metal semispaces whose dielectric property is described by the plasma model. Application  to the Casimir effect in Randall-Sundrum spacetime  is briefly discussed.
\end{abstract}

\maketitle
\section{Introduction}

The Casimir effect is an interesting phenomena that is due to the vacuum fluctuations of quantum fields. The Casimir effect of massless scalar field, electromagnetic field (massless vector fields) and massive scalar field have been extensively studied (see e.g. \cite{18,19,21,20}). In contrast, massive vector field  has not been much considered, mostly due to its substantial complication compared to the massless case. Nevertheless, massive vector field plays an important role in the study of physics with extra dimensions. To study a quantum field in a spacetime with extra dimensions, one can use the Kaluza-Klein decomposition to decompose the quantum field to an infinite tower of massive fields in four dimensions. This approach has been intrinsically used in a number of works to study the Casimir effect of scalar field or spinor field in spacetime with extra dimensions such as Kaluza-Klein spacetime and Randall-Sundrum spacetime. For electromagnetic field, thanks to the work of Barton and Dombey \cite{1}, the Casimir effect between two parallel perfectly conducting plates in Kaluza-Klein spacetime of the form $M^4\times S^1$ and in Randall-Sundrum model have been computed respectively in \cite{13} and \cite{11}.

The Casimir effect of a massive vector field was first considered in \cite{1}, and an explicit formula for the Casimir energy has been derived for two perfectly conducting plates of finite thicknesses. Asymptotic behaviors of the massive corrections were computed. Unlike the electromagnetic field, a massive vector field has three polarizations. Besides two transverse polarizations that correspond to the two polarizations of an electromagnetic field, a massive vector field also have longitudinal modes and these latter modes  can penetrate through perfectly conducting objects. In a recent work \cite{4}, we generalized the work \cite{1} to  magnetodielectric slabs. We found that for  Casimir effect of massive vector field on general magnetodielectric slabs, one cannot separate the contribution from the transverse magnetic modes from the contribution  of longitudinal modes, and we wound up with a very complicated formula for the Casimir interaction energy. The TM contribution, which is a combination of the contributions from the transverse modes and the longitudinal modes, is expressed in terms of log determinant of a four by four matrix. It is almost impossible to study such a formula analytically.

To study the electromagnetic Casimir effect of two semi-infinite dielectric slabs, Lifshitz \cite{22} has derived a formula which was later named after him. The first goal of this article is to derive a formula that looks more similar to the Lifshitz formula for the  Casimir interaction between two semi-infinite magnetodielectric slabs due to the fluctuation of a massive vector field. The second objective is to consider the deviation from a perfect conductor. As in the works \cite{12,3}, we model the two semi-infinite slabs as real metals whose dielectric property is   described by plasma model:
$$\vep(\omega)=\vep_0\left(1-\frac{\omega_p^2}{\omega^2}\right),$$
where $\vep_0$ is the vacuum permittivity and $\omega_p$ is the effective plasma frequency. We expand the generalized Lifshitz formula perturbatively in the dimensionless variable $$\alpha=\frac{c}{\omega_p d},$$ where $d$ is the distance between the slabs. $\alpha=0$ gives the perfect conductor limit, and a small $\alpha$ amounts to finite conductivity corrections. For the zero and first order in $\alpha$, we study the asymptotic behaviors of the Casimir energy and the Casimir force in various regions: large mass high temperature, large mass low temperature, small mass high temperature and small mass low temperature. In the last section, we also briefly discuss the application of these results to the Casimir effect of real metals in Randall-Sundrum spacetime.

\section{Massive vector field}
In this section, we recall some basic facts about massive vector fields that are required for the study of Casimir effect. For more details, one can refer to \cite{1,4}.
\subsection{Proca equations}
Define the electric field $\mathbf{E}$ and the magnetic field $\mathbf{B}$ by \begin{align}\label{eq1_5}\mathbf{E}=-\frac{\pa\mathbf{A}}{\pa t}-\nabla\phi,\hspace{1cm}\mathbf{B}=\nabla\times \mathbf{A}.\end{align}
The Proca equations for a massive vector field propagating in a medium with permittivity $\vep$ and permeability $\mu$ are \cite{1, 4, 5}:
\begin{equation}\begin{split}\label{eq1_1}
&\nabla\cdot\mathbf{B}=0,\\
&\nabla\times\mathbf{E}+\frac{\pa\mathbf{B}}{\pa t}=0, \\
&\nabla\cdot\mathbf{D}+\frac{m^2 }{\mu\hbar^2}\phi=  \rho_f,  \\
&\nabla\times \mathbf{H}-\frac{\pa\mathbf{D}}{\pa t}+ \frac{m^2c^2}{ \mu\hbar^2}\mathbf{A}=\mathbf{J}_f. \end{split}
\end{equation}
The first two are the well-known Maxwell's equations which are automatically satisfied because of \eqref{eq1_5}.
   The   continuity equation
\begin{align}\label{eq6_24_1}
\frac{\pa\rho_f}{\pa t}+\nabla\cdot\mathbf{J}_f=0
\end{align}  implies that
 the Lorentz condition
\begin{align}\label{eq1_6}
\frac{1}{c^2}\frac{\pa\phi}{\pa t}+ \nabla\cdot\mathbf{A}=0
\end{align}has to be satisfied.

 Let
\begin{equation*}
\begin{split}
\phi(\mathbf{x},t)=&\int_{-\infty}^{\infty}d\omega\phi(\mathbf{x},\omega)e^{-i\omega t},\\
\mathbf{A}(\mathbf{x},t)=&\int_{-\infty}^{\infty}d\omega \mathbf{A}(\mathbf{x},\omega)e^{-i\omega t},
\end{split}
\end{equation*}
and assume the linear relations
\begin{equation*}
\begin{split}
&\mathbf{D}(\mathbf{x},\omega)=\vep(\omega)\mathbf{E}(\mathbf{x},\omega),\\
&\mathbf{H}(\mathbf{x},\omega)=\frac{1}{\mu(\omega)}\mathbf{B}(\mathbf{x},\omega).
\end{split}
\end{equation*}
When there are no free charges (i.e. $\rho_f=0$) and free current (i.e. $\mathbf{J}_f=\mathbf{0}$),  the Proca equations  \eqref{eq1_1} are equivalent to the following two equations for $\phi(\mathbf{x},\omega)$ and $\mathbf{A}(\mathbf{x},\omega)$:
\begin{equation}\label{eq1_7}\begin{split}
\left(-\nabla^2-\frac{\omega^2}{c^2} +\frac{m^2 }{\varepsilon\mu\hbar^2}\right)\phi(\mathbf{x},\omega)=&0,
 \\
\nabla\times\nabla\times\mathbf{A}-\vep\mu c^2\nabla(\nabla\cdot\mathbf{A})-\left(\vep\mu\omega^2-\frac{m^2c^2}{\hbar^2}\right)\mathbf{A}=&\mathbf{0}.\end{split}
\end{equation}The Lorentz condition \eqref{eq1_6}
becomes
\begin{equation}\label{eq5_17_1}
\phi=-\frac{ic^2}{\omega}\nabla\cdot\mathbf{A}.
\end{equation}

The solutions of these equations can be divided into two types:

\begin{enumerate}
\item[$\bullet$] \textbf{Transverse waves.} These are waves with $\nabla\cdot \mathbf{A}=0$, $\phi=0$ and
\begin{equation*}
\nabla\times\nabla\times\mathbf{A}- \left(\vep\mu\omega^2-\frac{m^2c^2}{\hbar^2}\right)\mathbf{A}=\mathbf{0}
\end{equation*}
To solve this, we look for a vector wave function $\mathbf{A}$ satisfying
$$\nabla\times\nabla\times\mathbf{A}=k_T^2\mathbf{A}.$$
Then $\omega$ satisfies the dispersion relation
$$k_T^2= \vep\mu\omega^2-\frac{m^2c^2}{\hbar^2}.$$
The solutions can be divided into two families $\mathbf{A}^{\text{TE}}$ and $\mathbf{A}^{\text{TM}}$, which in the massless limit correspond to the   TE (transverse electric) and TM (transverse magnetic) waves of electromagnetic fields.

\vspace{0.5cm}
\item[$\bullet$] \textbf{Longitudinal waves}. These are waves with $\nabla\times \mathbf{A}=\mathbf{0}$. The equations \eqref{eq1_7} become
\begin{align*}
&\left(-\nabla^2-\frac{\omega^2}{c^2} +\frac{m^2 }{\varepsilon\mu\hbar^2}\right)\phi(\mathbf{x},\omega)=0,\\
&-\vep\mu c^2\nabla(\nabla\cdot\mathbf{A})-\left(\vep\mu\omega^2-\frac{m^2c^2}{\hbar^2}\right)\mathbf{A}=\mathbf{0}.
\end{align*}
To solve this system, one first find a scalar function $\phi$ satisfying
$$-\nabla^2\phi=k_L^2\phi.$$ Then $\mathbf{A}$ is given by
\begin{equation*}
\begin{split}
\mathbf{A}=-\frac{i\omega}{k_L^2c^2}\nabla\phi,
\end{split}
\end{equation*}and the dispersion relation is
\begin{equation*}
k_L^2=\frac{\omega^2}{c^2}-\frac{m^2 }{\varepsilon\mu\hbar^2}.
\end{equation*}
In this case, $\mathbf{B}=\mathbf{0}$ and
\begin{equation*}
\mathbf{E}=\frac{m^2}{\hbar^2\vep\mu k_L^2}\nabla\phi.
\end{equation*}

\end{enumerate}

\subsection{Plane waves solutions}
In this article, we only consider plane waves. Choosing $z$ as the distinctive direction,
the plane waves are parametrized by $\mathbf{k}_{\perp}=(k_x,k_y)\in\mathbb{R}^2$. The transverse TE and TM waves are:
\begin{equation*}
\begin{split}
\mathbf{A}^{\text{TE}, \,\substack{\text{reg}\\\text{out}}}_{\mathbf{k}_{\perp}}(\mathbf{x},\omega)=&\frac{1}{k_{\perp}}\nabla\times\left(e^{ik_xx+ik_yy\mp i\sqrt{k_T^2-k_{\perp}^2}z}\mathbf{e}_z\right)\\
=&\left(\frac{ik_y}{k_{\perp}}\mathbf{e}_x-\frac{ik_x}{k_{\perp}}\mathbf{e}_y\right)e^{ik_xx+ik_yy\mp i\sqrt{k_T^2-k_{\perp}^2}z},\\
\mathbf{A}^{\text{TM},\,\substack{\text{reg}\\\text{out}}}_{\mathbf{k}_{\perp}}(\mathbf{x},\omega)=&\frac{1}{k_Tk_{\perp}}\nabla\times\nabla\times\left(e^{ik_xx+ik_yy\mp i\sqrt{k_T^2-k_{\perp}^2}z}\mathbf{e}_z\right)\\
=&
\left(\pm\frac{ k_x\sqrt{k_T^2-k_{\perp}^2}}{k_Tk_{\perp}}\mathbf{e}_x \pm \frac{k_y\sqrt{k_T^2-k_{\perp}^2}}{k_Tk_{\perp}}\mathbf{e}_y+\frac{k_{\perp}}{k_T}\mathbf{e}_z\right)e^{ik_xx+ik_yy\mp i\sqrt{k_T^2-k_{\perp}^2}z}.
\end{split}
\end{equation*}
The longitudinal waves are given by
\begin{equation*}
\begin{split}
\phi^{\text{L}, \,\substack{\text{reg}\\\text{out}}}_{\mathbf{k}_{\perp}}(\mathbf{x},\omega)=&\frac{c^2 k_{L}^2}{k_{\perp}\omega}e^{ik_xx+ik_yy\mp i\sqrt{k_L^2-k_{\perp}^2}z},\\
\mathbf{A}^{\text{L}, \,\substack{\text{reg}\\\text{out}}}_{\mathbf{k}_{\perp}}(\mathbf{x},\omega)=&
\left( \frac{ k_x }{k_{\perp}}\mathbf{e}_x+\frac{ k_y }{k_{\perp} }\mathbf{e}_y\mp \frac{   \sqrt{k_L^2-k_{\perp}^2}}{k_{\perp} }\mathbf{e}_z\right)e^{ik_xx+ik_yy\mp i\sqrt{k_L^2-k_{\perp}^2}z}.
\end{split}
\end{equation*}Here reg and out stand  for regular waves and outgoing waves respectively.

\section{Finite temperature Casimir interaction of two magnetodielectric slabs separated by a magnetodielectric medium}
\begin{figure}[h]
\epsfxsize=0.5\linewidth \epsffile{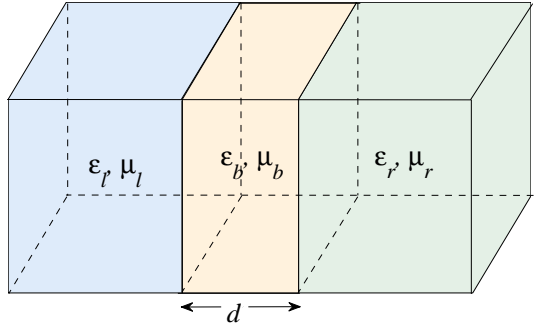} \caption{\label{f1}Two parallel magnetodielectric slabs separated by a magnetodielectric medium. }\end{figure}
 Consider two parallel slabs, each of area $A$, occupying the regions $-L<z<0$ and $d<z<L$, with permittivities $\vep_l,   \vep_r$ and permeabilities $\mu_l,   \mu_r$ respectively. The medium between the slabs is assumed to have permittivity $\vep_b$ and permeability $\mu_b$ (see Fig. \ref{f1}). As discussed in \cite{1,4}, the
  boundary conditions of a massive vector field are the continuities of $\displaystyle  \mathbf{A}, \mu^{-1}\mathbf{B}_{\parallel}$ and the continuity of either $\phi$ or $\pa_n\mathbf{A}_n$, where $n$ is a unit normal vector to the boundary.

  In the following, we compute  the contribution to the Casimir free interaction energy from the transverse TE modes, and from the combination of transverse TM modes and longitudinal modes. We denote the former as TE contribution, and the latter as TM contribution.

\subsection{TE contribution} For TE contribution to the Casimir free interaction energy, $\phi(\mathbf{x},t)=0$ and
\begin{equation*}
\begin{split}
\mathbf{A}(\mathbf{x},t)=A\int_{-\infty}^{\infty}d\omega\int_{-\infty}^{\infty}\frac{dk_x}{2\pi}\int_{-\infty}^{\infty}\frac{dk_y}{2\pi}
\left(\mathfrak{A}_*(\mathbf{k}_{\perp})\mathbf{A}_{\mathbf{k}_{\perp},*}^{\text{TE, reg}}(\mathbf{x},\omega)+\mathfrak{B}_*(\mathbf{k}_{\perp})\mathbf{A}_{\mathbf{k}_{\perp},*}^{\text{TE, out}}(\mathbf{x},\omega)\right)e^{-i\omega t}.
\end{split}
\end{equation*}Here $*=l$ when $-L<z<0$, $*=b$ when $0<z<d$ and $*=r$ when $d<z<L$,
\begin{equation*}
\mathbf{A}^{\text{TE}, \,\substack{\text{reg}\\\text{out}}}_{\mathbf{k}_{\perp},*}(\mathbf{x},\omega)
=\left(\frac{ik_y}{k_{\perp}}\mathbf{e}_x-\frac{ik_x}{k_{\perp}}\mathbf{e}_y\right)e^{ik_xx+ik_yy\mp ip_{T,*}z},
\end{equation*}
$p_{T,*}=\sqrt{k_{T,*}^2-k_{\perp}^2}$. Taking the limit $L\rightarrow \infty$ amounts to setting $\mathfrak{B}_l=0$ and $\mathfrak{A}_r=0$.
The continuities of $\mathbf{A}$ and $\mu^{-1}\mathbf{B}_{\parallel}$ then give
\begin{equation*}
\begin{split}
&\mathfrak{A}_l =\mathfrak{A}_b+\mathfrak{B}_b,\\
& \frac{p_{T,l}}{\mu_l}\mathfrak{A}_l=\frac{p_{T,b}}{\mu_b}\left(\mathfrak{A}_b-\mathfrak{B}_b\right),\\
&\mathfrak{B}_re^{ip_{T,r}d}=\mathfrak{A}_b e^{-ip_{T,b}d}+\mathfrak{B}_be^{ip_{T,b}d},\\
&-\frac{p_{T,r}}{\mu_r} \mathfrak{B}_re^{ip_{T,r}d}=\frac{p_{T,b}}{\mu_b}\left(\mathfrak{A}_be^{-ip_{T,b}d}-\mathfrak{B}_be^{ip_{T,b}d}\right). \\
\end{split}
\end{equation*}
Eliminating $\mathfrak{A}_l$ and $\mathfrak{B}_r$ from these equations, we obtain
\begin{equation}\label{eq5_31_1}
\begin{split}
&\left(p_{T,b}\mu_l+p_{T,l}\mu_b\right)\mathfrak{B}_b=\left(p_{T,b}\mu_l-p_{T,l}\mu_b\right)\mathfrak{A}_b,\\
&\left(p_{T,b}\mu_r-p_{T,r}\mu_b\right)\mathfrak{B}_be^{ip_{T,b}d}=\left(p_{T,b}\mu_r+p_{T,r}\mu_b\right)\mathfrak{A}_be^{-ip_{T,b}d}.
\end{split}
\end{equation}
To have   nontrivial solutions for $(\mathfrak{A}_b,\mathfrak{B}_b)$, we find that the eigenfrequencies $\omega$ should satisfy the equation
\begin{equation*}
\left(p_{T,b}\mu_l+p_{T,l}\mu_b\right)\left(p_{T,b}\mu_r+p_{T,r}\mu_b\right) e^{-ip_{T,b}d}-\left(p_{T,b}\mu_l-p_{T,l}\mu_b\right)\left(p_{T,b}\mu_r-p_{T,r}\mu_b\right)e^{ip_{T,b}d}=0.
\end{equation*}
Factoring out
$$\left(p_{T,b}\mu_l+p_{T,l}\mu_b\right)\left(p_{T,b}\mu_r+p_{T,r}\mu_b\right) e^{-ip_{T,b}d}$$ which can be attributed to the self-energies of the individual slabs, we obtain
\begin{equation}\label{eq5_31_2}1-\frac{\left(p_{T,b}\mu_l-p_{T,l}\mu_b\right)\left(p_{T,b}\mu_r-p_{T,r}\mu_b\right)}{\left(p_{T,b}\mu_l+p_{T,l}\mu_b\right)\left(p_{T,b}\mu_r+p_{T,r}\mu_b\right) }e^{2ip_{T,b}d}=0.
\end{equation}
Alternatively, we can derive this
by writing \eqref{eq5_31_1}
in the form
\begin{equation*}
\begin{split}
& \mathfrak{B}_b=-R_{l}^{\text{TE}}\mathfrak{A}_b,\\
&-R_r^{\text{TE}}\mathfrak{B}_be^{ip_{T,b}d}= \mathfrak{A}_be^{-ip_{T,b}d},
\end{split}
\end{equation*}
where
\begin{equation*}
R_*^{\text{TE}}=\frac{p_{T,*} \mu_b -p_{T,b} \mu_* }{p_{T,*} \mu_b +p_{T,b} \mu_* },\quad *=l\;\text{or}\;r.
\end{equation*}
Then
\begin{equation*}
\mathfrak{B}_b=-R_{l}^{\text{TE}}\mathfrak{A}_b=R_{l}^{\text{TE}}R_r^{\text{TE}}e^{2ip_{T,b}d}\mathfrak{B}_b.
\end{equation*}
From this, we find that there are nontrivial solutions for $\mathfrak{B}_b$ if and only if $\omega$ satisfies the relation \eqref{eq5_31_2}.
Hence, by using the standard contour integration technique and the Matsubara formalism, we find that the  TE contribution to the Casimir  free interaction energy is
\begin{equation}\label{eq6_15_2}
E_{\text{Cas}}^{\text{TE}}=\frac{k_BT A}{2\pi }\sum_{n=0}^{\infty}\!'\int_0^{\infty}dk_{\perp}k_{\perp} \ln\left(1-R^{\text{TE}}_l(i\xi_n)R^{\text{TE}}_r(i\xi_n)e^{-2q_{T,b}(\xi_n)d}\right),
\end{equation}
where $$\xi_n=\frac{2\pi n k_B T}{\hbar}$$ are the Matsubara frequencies,
\begin{equation}\label{eq6_15_3}
R_*^{\text{TE}}(i\xi)=\frac{q_{T,*}(\xi) \mu_b(i\xi) -q_{T,b}(\xi) \mu_* (i\xi)}{q_{T,*}(\xi) \mu_b (i\xi)+q_{T,b} (\xi)\mu_*(i\xi) },\quad *=l\;\text{or}\;r,
\end{equation}
\begin{equation*}
q_{T,*}(\xi) =\sqrt{\vep_*(i\xi)\mu_*(i\xi)\xi^2+\frac{m^2c^2}{\hbar^2}+k_{\perp}^2},\quad *=l, b\;\text{or}\;r.
\end{equation*}

\subsection{TM contribution}
  For the TM contribution to the Casimir energy, which comes from the transverse TM modes and the longitudinal modes, let
\begin{equation*}
\begin{split}
\phi(\mathbf{x},t)=A\int_{-\infty}^{\infty}d\omega\int_{-\infty}^{\infty}\frac{dk_x}{2\pi}\int_{-\infty}^{\infty}\frac{dk_y}{2\pi}\left(\mathfrak{E}_*(\mathbf{k}_{\perp})\phi_{\mathbf{k}_{\perp},*}^{\text{L,\,reg}}(\mathbf{x},\omega)+
\mathfrak{F}_*(\mathbf{k}_{\perp})\phi_{\mathbf{k}_{\perp},*}^{\text{L,\,out}}(\mathbf{x},\omega)\right)e^{-i\omega t},
\end{split}
\end{equation*}and
  \begin{equation*}
\begin{split}
\mathbf{A}(\mathbf{x},t)=&A\int_{-\infty}^{\infty}d\omega\int_{-\infty}^{\infty}\frac{dk_x}{2\pi}\int_{-\infty}^{\infty}\frac{dk_y}{2\pi}\\&\times \left(
\mathfrak{C}_*(\mathbf{k}_{\perp})\mathbf{A}_{\mathbf{k}_{\perp},*}^{\text{TM,\,reg}}(\mathbf{x},\omega)+
\mathfrak{D}_*(\mathbf{k}_{\perp})\mathbf{A}_{\mathbf{k}_{\perp},*}^{\text{TM,\,out}}(\mathbf{x},\omega)+
\mathfrak{E}_*(\mathbf{k}_{\perp})\mathbf{A}_{\mathbf{k}_{\perp},*}^{\text{L,\,reg}}(\mathbf{x},\omega)+
\mathfrak{F}_*(\mathbf{k}_{\perp})\mathbf{A}_{\mathbf{k}_{\perp},*}^{\text{L,\,out}}(\mathbf{x},\omega)\right)e^{-i\omega t},
\end{split}
\end{equation*}
where $*=l$ when $-L<z<0$, $*=b$ when $0<z<d$ and $*=r$ when $d<z<L$;
\begin{equation*}
\begin{split}
\phi^{\text{L}, \,\substack{\text{reg}\\\text{out}}}_{\mathbf{k}_{\perp},*}(\mathbf{x},\omega)=&\frac{c^2 k_{L,*}^2}{k_{\perp}\omega}e^{ik_xx+ik_yy\mp ip_{L,*}z},
\end{split}
\end{equation*}
  \begin{equation*}
\begin{split}
\mathbf{A}^{\text{TM},\,\substack{\text{reg}\\\text{out}}}_{\mathbf{k}_{\perp},*}(\mathbf{x},\omega)
=&
\left(\pm\frac{ k_xp_{T,*} }{k_{T,*}k_{\perp}}\mathbf{e}_x \pm \frac{k_yp_{T,*}}{k_{T,*}k_{\perp}}\mathbf{e}_y+\frac{k_{\perp}}{k_{T,*}}\mathbf{e}_z\right)e^{ik_xx+ik_yy\mp ip_{T,*}z},
\end{split}
\end{equation*}
\begin{equation*}
\begin{split}
\mathbf{A}^{\text{L}, \,\substack{\text{reg}\\\text{out}}}_{\mathbf{k}_{\perp},*}(\mathbf{x},\omega)=&
\left( \frac{ k_x }{k_{\perp}}\mathbf{e}_x+\frac{ k_y }{k_{\perp}}\mathbf{e}_y\mp \frac{    p_{L,*}}{k_{\perp}}\mathbf{e}_z\right)e^{ik_xx+ik_yy\mp ip_{L,*}z}.
\end{split}
\end{equation*}
Here
$$p_{L,*}=\sqrt{k_{L,*}^2-k_{\perp}^2}.$$
Taking the limit $L\rightarrow \infty$ amounts to setting $\mathfrak{D}_l=0, \mathfrak{F}_l=0, \mathfrak{C}_r=0$ and $\mathfrak{E}_r=0$.
The continuities of $\mathbf{A}, \phi$ and $\mu^{-1}\mathbf{B}_{\parallel}$ then give the following two sets of relations:
\begin{equation*}
\begin{split}
&k_{L,l}^2\mathfrak{E}_l=k_{L,b}^2\left(\mathfrak{E}_b+\mathfrak{F}_b\right),\\
&\frac{k_{T,l}}{\mu_l}\mathfrak{C}_l=\frac{k_{T,b}}{\mu_b}(\mathfrak{C}_b+\mathfrak{D}_b),\\
&\frac{p_{T,l}}{k_{T,l} }\mathfrak{C}_l+ \mathfrak{E}_l= \frac{p_{T,b}}{k_{T,b} }(\mathfrak{C}_b-\mathfrak{D}_b)+ (\mathfrak{E}_b+\mathfrak{F}_b),\\
&\frac{k_{\perp}}{k_{T,l} }\mathfrak{C}_l- \frac{  p_{L,l}}{k_{\perp}}\mathfrak{E}_l= \frac{k_{\perp}}{k_{T,b}}(\mathfrak{C}_b+\mathfrak{D}_b)- \frac{ p_{L,b}}{k_{\perp}}(\mathfrak{E}_b-\mathfrak{F}_b),\hspace{6cm}\\
\end{split}\end{equation*}
\begin{equation*}
\begin{split}
&k_{L,r}^2\mathfrak{F}_re^{ip_{L,r}d}=k_{L,b}^2\left(\mathfrak{E}_be^{-ip_{L,b}d}+\mathfrak{F}_b e^{ip_{L,b}d}\right),\\
&\frac{k_{T,r}}{\mu_r}\mathfrak{D}_r e^{ip_{T,r}d}=\frac{k_{T,b}}{\mu_b}(\mathfrak{C}_be^{-ip_{T,b}d}+\mathfrak{D}_be^{ip_{T,b}d}),\\
&-\frac{p_{T,r}}{k_{T,r} }\mathfrak{D}_re^{ip_{T,r}d}+ \mathfrak{F}_re^{ip_{L,r}d}= \frac{p_{T,b}}{k_{T,b} }(\mathfrak{C}_b
e^{-ip_{T,b}d}-\mathfrak{D}_be^{ip_{T,b}d})+  (\mathfrak{E}_be^{-ip_{L,b}d}+\mathfrak{F}_be^{ip_{L,b}d}),\\
&\frac{k_{\perp}}{k_{T,r} }\mathfrak{D}_re^{ip_{T,r}d}+\frac{ p_{L,r}}{k_{\perp}}\mathfrak{F}_re^{ip_{L,r}d}= \frac{k_{\perp}}{k_{T,b}}(\mathfrak{C}_be^{-ip_{T,b}d}+\mathfrak{D}_b
e^{ip_{T,b}d})- \frac{  p_{L,b}}{k_{\perp}}(\mathfrak{E}_be^{-ip_{L,b}d}-\mathfrak{F}_be^{ip_{L,b}d}).
\end{split}\end{equation*}
Eliminating $\mathfrak{C}_l, \mathfrak{E}_l, \mathfrak{D}_r, \mathfrak{F}_r$ give relations of the form
\begin{equation*}
\begin{split}
&\begin{pmatrix} \mathfrak{D}_b\\\mathfrak{F}_b\end{pmatrix}=-\boldsymbol{R}^{\text{TM}}_l\begin{pmatrix} \mathfrak{C}_b\\\mathfrak{E}_b\end{pmatrix},\\
&\begin{pmatrix} e^{-ip_{T,b}d} & 0\\ 0&e^{-ip_{L,b}d}\end{pmatrix}\begin{pmatrix} \mathfrak{C}_b\\\mathfrak{E}_b\end{pmatrix}=-\boldsymbol{R}^{\text{TM}}_r\begin{pmatrix} e^{ip_{T,b}d} & 0\\ 0&e^{ip_{L,b}d}\end{pmatrix}\begin{pmatrix} \mathfrak{D}_b\\\mathfrak{F}_b\end{pmatrix},
\end{split}
\end{equation*}
where $\boldsymbol{R}^{\text{TM}}_l$ and $\boldsymbol{R}^{\text{TM}}_r$ are $2\times 2$ matrices.
Hence, the TM contribution to the Casimir free interaction energy is
\begin{equation*}
E_{\text{Cas}}^{\text{TE}}=\frac{k_B T A}{2\pi}\sum_{n=0}^{\infty}\!'\int_0^{\infty}dk_{\perp}k_{\perp}  \ln\det\left(1-\boldsymbol{R}^{\text{TM}}_l
(i\xi_n)\boldsymbol{U}(i\xi_n)\boldsymbol{R}^{\text{TM}}_r(i\xi_n)\boldsymbol{U}(i\xi_n)\right),
\end{equation*}
where
\begin{equation*}
\begin{split}
\boldsymbol{U}=\begin{pmatrix}
e^{-q_{T,b}d} & 0\\ 0 & e^{-q_{L,b}d}
\end{pmatrix},\hspace{1cm}\boldsymbol{R}_*^{\text{TM}}=\begin{pmatrix}R^{\text{TM}}_{*,11} &R^{\text{TM}}_{*,12}\\
R^{\text{TM}}_{*,21}&R^{\text{TM}}_{*,22}\end{pmatrix},\quad *=l\,\text{or}\,r.
\end{split}
\end{equation*}
For $*=l$ or $r$,
\begin{equation*}
\begin{split}
\Delta_*=&\left(q_{T,*}\mu_*\kappa_{T,b}^2+q_{T,b}\mu_b\kappa_{T,*}^2\right) \left(q_{L,b}\kappa_{L,*}^2+q_{L,*}\kappa_{L,b}^2\right)
+\frac{m^2c^2k_{\perp}^2}{\hbar^2}\frac{\left(n_b^2-n_*^2\right)}{n_b^2n_*^2}\left(\kappa_{T,b}^2\mu_*-\kappa_{T,*}^2\mu_b\right),\\
R^{\text{TM}}_{*,11}=& 1- \frac{2q_{T,b}\kappa_{T,*}^2\mu_b\left(q_{L,b}\kappa_{L,*}^2+q_{L,*}\kappa_{L,b}^2\right)}{\Delta_*},\\
R_{*,12}^{\text{TM}}=& -\sigma_*\frac{2   \kappa_{T,*}^2q_{L,b}\kappa_{T,b}\mu_b}{ \Delta_* } \frac{m^2c^2}{\hbar^2}\frac{n_b^2-n_*^2}{n_b^2n_*^2},\\
R_{*,21}^{\text{TM}}= &\sigma_*\frac{2q_{T,b}\kappa_{L,*}^2k_{\perp}^2}{\kappa_{T,b}\Delta_*}(\kappa_{T,b}^2\mu_*-\kappa_{T,*}^2\mu_b),\\
R_{*,22}^{\text{TM}}=& 1-\frac{2\kappa_{L,*}^2q_{L,b}\left(q_{T,*}\mu_*\kappa_{T,b}^2+q_{T,b}\mu_b\kappa_{T,*}^2\right)}{\Delta_*};
\end{split}
\end{equation*}
$\sigma_*=1$ for $*=l$ and $\sigma_*=-1$ for $*=r$.
For $*=l,b,r$,
\begin{equation*}
\begin{split}
\kappa_{T,*}=&\sqrt{\vep_*(i\xi)\mu_*(i\xi)\xi^2+\frac{m^2c^2}{\hbar^2}}, \hspace{1cm}
q_{T,*}= \sqrt{\kappa_{T,*}^2+k_{\perp}^2}, \\
\kappa_{L,*}=&\sqrt{\frac{\xi^2}{c^2}+\frac{m^2 }{\hbar^2\vep_*(i\xi)\mu_*(i\xi)}}, \hspace{1cm}
q_{L,*}=\sqrt{\kappa_{L,*}^2+k_{\perp}^2}.
\end{split}\end{equation*}
This presentation of the TM contribution has substantially simplifies the presentation given in \cite{4}. Moreover, it has been cast in the form of TGTG formula \cite{6,7}, which is the general presentation for the Casimir interaction energy between two objects.

\section{Massless limit}
In this section, we consider the limiting case when the mass is zero.
When $m=0$, it follows immediately from \eqref{eq6_15_2} that
the  TE contribution to the Casimir free interaction energy is given by the same formula \eqref{eq6_15_2}, with $R_*^{\text{TE}}$ given by \eqref{eq6_15_3}, but $q_{T,*}$ is reduced to
\begin{equation*}
q_{T,*}(\xi)=\sqrt{\vep_*(i\xi)\mu_*(i\xi)\xi^2+k_{\perp}^2},\quad *=l, b\;\text{or}\;r.
\end{equation*}
For the TM contribution,
obviously, when $m=0$, $$R_{l,12}^{\text{TM}}=R_{r,12}^{\text{TM}}=0.$$On the other hand, when $m=0$,
\begin{equation*}
k_{L,l}=k_{L,r}=k_{L,b}=\frac{\xi}{c},\hspace{1cm} q_{L,l}=q_{L,r}=q_{L,b}=\sqrt{\frac{\xi^2}{c^2}+k_{\perp}^2}.
\end{equation*}Therefore,
\begin{equation*}
R_{l,22}^{\text{TM}}=R_{r,22}^{\text{TM}}=0.
\end{equation*}
Hence, the  TM contribution to the Casimir free interaction energy reduce to
\begin{equation*}\begin{split}
E_{\text{Cas}}^{\text{TM}}=&\frac{k_BT A}{2\pi}\sum_{n=0}^{\infty}\!'\int_0^{\infty}dk_{\perp}k_{\perp}  \ln\left(1-R_{l,11}^{\text{TM}}(i\xi_n)R_{r,11}^{\text{TM}}(i\xi_n)e^{-2q_{T,b}(\xi_n)d}\right),
\end{split}
\end{equation*}
where
\begin{equation*}
\begin{split}
R_{*,11}^{\text{TM}}=&1-\frac{2q_{T,b}\vep_*\mu_*\mu_b}{q_{T,*}\mu_*\vep_b\mu_b+q_{T,b}\mu_b\vep_*\mu_*}\\
=&\frac{q_{T,*}\vep_b-q_{T,b}\vep_*}{q_{T,*}\vep_b+q_{T,b}\vep_*}.
\end{split}
\end{equation*}
These recovers the Lifshitz's formula \cite{22} for the electromagnetic Casimir free interaction energy between two dielectric slabs.

\section{The limits of parallel perfect conductors separated by vacuum}

In this section, we consider the case where two perfectly conducting plates are separated by vacuum. This can be achieved by taking the limit $\vep_l,\vep_r\rightarrow \infty$ and setting $\vep_b=\vep_0$, $\mu_b=\mu_0$. In this case, we find that
\begin{equation*}
\begin{split}
\kappa_{T,b}=&\kappa_{L,b}=\sqrt{\frac{\xi^2}{c^2}+\frac{m^2c^2}{\hbar^2}}:=\kappa_b,\\
q_{T,b}=&q_{L,b}=\sqrt{\frac{\xi^2}{c^2}+\frac{m^2c^2}{\hbar^2}+k_{\perp}^2}:=q_b,
\end{split}
\end{equation*}
for $*=l$ or $r$,
\begin{equation}\label{eq6_18_1}
\begin{split}
&\kappa_{T,*}, q_{T, *}\rightarrow\infty,\\
&\kappa_{L,*}\rightarrow \frac{\xi}{c}:=\kappa_0,\\
&q_{L,*}\rightarrow\sqrt{\frac{\xi^2}{c^2}+k_{\perp}^2}:=q_0,\\
&R_{*,11}^{\text{TM}}\rightarrow \frac{-q_b(q_0\kappa_b^2+q_b\kappa_0^2)+\frac{m^2c^2k_{\perp}^2}{\hbar^2}}{q_b(q_0\kappa_b^2+q_b\kappa_0^2)+\frac{m^2c^2k_{\perp}^2}{\hbar^2}}
=\frac{-q_0(q_b^2+k_{\perp}^2)+2q_bk_{\perp}^2}{q_0(q_b^2-k_{\perp}^2)},\\
&R_{*,12}^{\text{TM}}\rightarrow \sigma_* \frac{2q_b\kappa_b\frac{m^2c^2}{\hbar^2}}{q_b(q_0\kappa_b^2+q_b\kappa_0^2)+\frac{m^2c^2k_{\perp}^2}{\hbar^2}}
=\sigma_*\frac{2q_b\kappa_b(q_b-q_0)}{q_0(q_b^2-k_{\perp}^2)},\\
&R_{*, 21}^{\text{TM}}\rightarrow -\sigma_* \frac{2q_b\kappa_0^2k_{\perp}^2}{\kappa_b\left[q_b(q_0\kappa_b^2+q_b\kappa_0^2)+\frac{m^2c^2k_{\perp}^2}{\hbar^2}\right]}
=-\sigma_* \frac{2q_b(q_0^2-k_{\perp}^2)k_{\perp}^2}{\kappa_bq_0(q_b+q_0)(q_b^2-k_{\perp}^2)},\\
&R_{*,22}^{\text{TM}}\rightarrow \frac{ q_b(q_0\kappa_b^2-q_b\kappa_0^2)+\frac{m^2c^2k_{\perp}^2}{\hbar^2}}{q_b(q_0\kappa_b^2+q_b\kappa_0^2)+\frac{m^2c^2k_{\perp}^2}{\hbar^2}}
=\frac{q_b-q_0}{q_b+q_0}\frac{q_0(q_b^2+k_{\perp}^2)+2q_bk_{\perp}^2}{q_0(q_b^2-k_{\perp}^2)}.
\end{split}
\end{equation}
One can then show that
\begin{equation}\label{eq6_18_9} R_*^{\text{TE}}\rightarrow 1,\end{equation}
and
\begin{equation}\label{eq6_18_10}
\begin{split}
\det\left(1-\boldsymbol{R}^{\text{TM}}_l
 \boldsymbol{U} \boldsymbol{R}^{\text{TM}}_r \boldsymbol{U} \right)\rightarrow \left(1-e^{-2q_b d}\right)\left(1-\Lambda^2 e^{-2q_b d}\right),
\end{split}
\end{equation}where
\begin{equation*}
\Lambda=\frac{q_b-q_0}{q_b+q_0}.
\end{equation*}
For the zero Matsubara frequency $\xi_0$, we have to use the Schwinger-DeRaad-Milton prescription \cite{15} where the limit $\vep_l,\vep_r\rightarrow\infty$ is taken before setting $\xi_0=0$.

From \eqref{eq6_18_9} and \eqref{eq6_18_10}, we find that in the perfect conductor limit, the Casimir free interaction energy is given by
\begin{equation*}\begin{split}
E_{\text{Cas}}  =&\frac{k_BT A}{2\pi}\sum_{n=0}^{\infty}\!'\int_0^{\infty}dk_{\perp}k_{\perp}  \ln\left[\left(1- e^{-2q_{ b}(\xi_n)d}\right)^2
\left(1-\left[\frac{q_b(\xi_n)-q_0(\xi_n)}{q_b(\xi_n)+q_0(\xi_n)}\right]^2 e^{-2q_{ b}(\xi_n)d}\right)\right].
\end{split}
\end{equation*}
In the language of \cite{1}, the factor $\displaystyle \left(1- e^{-2q_{ b}(\xi_n)d}\right)^2$ gives two discrete mode contributions, whereas the factor
$$\left(1-\left[\frac{q_b(\xi_n)-q_0(\xi_n)}{q_b(\xi_n)+q_0(\xi_n)}\right]^2 e^{-2q_{ b}(\xi_n)d}\right)$$
gives a continuous mode contribution.
\section{Asymptotic behavior for real metals described by plasma model}
In this section, we consider the case where the two semi-infinite slabs are metals with permittivities described by plasma model:
\begin{equation*}
\begin{split}
\vep_l(i\xi)=\vep_r(i\xi)=\vep(i\xi)=&\vep_0\left(1+\frac{\omega_{p}^2}{\xi^2}\right),\\
\end{split}
\end{equation*}and permeabilities $\mu_l=\mu_r=\mu_0$.
Here $\omega_p$ is the effective plasma frequency.

The medium between the metals is assumed to be vacuum, i.e., $\vep_b=\vep_0$, $\mu_b=\mu_0$.
Let
\begin{equation*}
\begin{split}
\lambda=&\frac{mc d}{\hbar},\hspace{1cm}
\alpha= \frac{c}{\omega_pd}=\frac{\alpha_p}{2\pi d},\hspace{1cm}
z_n= \frac{\xi_nd}{c}.
\end{split}
\end{equation*}
Then
\begin{equation*}
\frac{\vep(i\xi_n)}{\vep_0}=1+\frac{1}{\alpha^2z_n^2}.
\end{equation*}
Making a change of variables
$$z=dq_b=d\sqrt{\frac{\xi_n^2}{c^2}+\frac{m^2c^2}{\hbar^2}+k_{\perp}^2},$$ we find that
the TE Casimir free interaction energy is given by
\begin{equation}\label{eq5_31_3}
\begin{split}
E_{\text{Cas}}^{\text{TE}}=&\frac{k_BT A}{2\pi d^2}\sum_{n=0}^{\infty}\!'\int_{\sqrt{z_n^2+\lambda^2}}^{\infty}dz z \ln\left(1-\left[R^{\text{TE}}\right]^2e^{-2z }\right),
\end{split}
\end{equation}where
\begin{equation*}
R^{\text{TE}}=\frac{\sqrt{1+\alpha^2z^2}-\alpha z }{\sqrt{1+\alpha^2z^2}+\alpha z }.
\end{equation*}
The TM  Casimir free interaction energy is more complicated:
\begin{equation}\label{eq5_31_4}
\begin{split}
E_{\text{Cas}}^{\text{TM}}=&\frac{k_BT A}{2\pi d^2}\sum_{n=0}^{\infty}\!'\int_{\sqrt{z_n^2+\lambda^2}}^{\infty}dz\,z  \ln\det\left(1- \boldsymbol{R}_l^{\text{TM}}
\boldsymbol{R}_r^{\text{TM}}e^{-2z}\right),
\end{split}\end{equation}
where
\begin{equation*}
\begin{split}
 \boldsymbol{R}_*^{\text{TM}}=\frac{1}{U}\begin{pmatrix}V_{11} &\sigma_*V_{12}\\
-\sigma_*V_{21}&V_{22}\end{pmatrix},
\end{split}
\end{equation*}
\begin{equation*}
\begin{split}
U=&\left(\alpha\sqrt{1+z^2\alpha^2}[z_n^2+\lambda^2]+z\left[1+\alpha^2(z_n^2+\lambda^2)\right]\right)
\left(zz_n^2\left[1+\alpha^2(z_n^2+\lambda^2)\right]+\sqrt{1+\alpha^2z_n^2}\sqrt{z^2-\lambda^2+\alpha^2z^2z_n^2}[z_n^2+\lambda^2]\right)\\
&+\lambda^2(z^2-z_n^2-\lambda^2),\\
V_{11}=& \left(\alpha\sqrt{1+z^2\alpha^2}[z_n^2+\lambda^2]-z\left[1+\alpha^2(z_n^2+\lambda^2)\right]\right)
\left(zz_n^2\left[1+\alpha^2(z_n^2+\lambda^2)\right]+\sqrt{1+\alpha^2z_n^2}\sqrt{z^2-\lambda^2+\alpha^2z^2z_n^2}[z_n^2+\lambda^2]\right)\\
&+\lambda^2(z^2-z_n^2-\lambda^2),\\
V_{12}=& 2\lambda^2z\sqrt{z_n^2+\lambda^2}\left(1+\alpha^2[z_n^2+\lambda^2]\right),\\
V_{21}= &\frac{2zz_n^2\left(1+\alpha^2[z_n^2+\lambda^2]\right)(z^2-z_n^2-\lambda^2)}{\sqrt{z_n^2+\lambda^2}},\\
V_{22}=& \left(\alpha\sqrt{1+z^2\alpha^2}[z_n^2+\lambda^2]+z\left[1+\alpha^2(z_n^2+\lambda^2)\right]\right)
\left(-zz_n^2\left[1+\alpha^2(z_n^2+\lambda^2)\right]+\sqrt{1+\alpha^2z_n^2}\sqrt{z^2-\lambda^2+\alpha^2z^2z_n^2}[z_n^2+\lambda^2]\right)\\
&+\lambda^2(z^2-z_n^2-\lambda^2).
\end{split}
\end{equation*}
A straightforward computation gives
\begin{equation*}
\begin{split}
\det\left(1- \boldsymbol{R}_l^{\text{TM}}
\boldsymbol{R}_r^{\text{TM}}e^{-2z}\right)=&1-\frac{V_{11}^2+V_{22}^2+2V_{12}V_{21}}{U^2}e^{-2z}+\frac{(V_{11}V_{22}+V_{12}V_{21})^2}{U^4}e^{-4z}\\
=&\left(1-T_+e^{-2z}\right)\left(1-T_-e^{-2z}\right),
\end{split}
\end{equation*}
where
\begin{equation*}
T_{\pm}=\frac{V_{11}^2+V_{22}^2+2V_{12}V_{21}\pm (V_{22}-V_{11})\sqrt{(V_{11}+V_{22})^2+4V_{12}V_{21}}}{2U^2}.
\end{equation*}
Hence,
\begin{equation}\label{eq5_31_5}
\begin{split}
E_{\text{Cas}}^{\text{TM}}=&\frac{k_BT A}{2\pi d^2}\sum_{n=0}^{\infty}\!'\int_{\sqrt{z_n^2+\lambda^2}}^{\infty}dz\,z  \ln\left[\left(1-T_+e^{-2z}\right)\left(1-T_-e^{-2z}\right)\right].
\end{split}\end{equation}

As in \cite{12,3}, we consider the asymptotic behaviors of the Casimir free interaction energy when $\alpha\ll 1$. $\alpha=0$ corresponds to the perfect conductor limit. Nonzero $\alpha$ corresponds to finite conductivity corrections. For aluminium, gold and copper plates, the plasma wavelength $\alpha_p$ are given respectively by 98nm, 132nm and 132nm \cite{16}. With separation between the plates $d$ in the range  100nm $=$ 0.1$\mu$m to 1$\mu$m, $\alpha<0.25$.

 The computations of the asymptotic behaviors are quite tedious and we leave them to the Appendix \ref{A1}. Here we discuss the results.

Let
$$\vartheta=\frac{2\pi k_B Td}{\hbar c}=\frac{\pi T}{T_{\text{eff}}},$$  where the effective temperature $$T_{\text{eff}}=\frac{\hbar c}{2k_B d} $$is defined in \cite{2}.

  In the large mass high temperature $\lambda,\vartheta\gg 1$ region, we find that
\begin{equation}
E_{\text{Cas}}\sim -\frac{3k_BTA}{8\pi d^2}\lambda e^{-2\lambda}\left(1-\frac{8\lambda }{3}\alpha+\ldots\right).
\end{equation}
Differentiating with respect to $d$, we find that the Casimir free interaction force
$$F_{\text{Cas}}=-\frac{\pa E_{\text{Cas}}}{\pa d}$$behaves like
\begin{equation}\label{eq6_18_2}
F_{\text{Cas}}\sim -\frac{3k_BTA}{4\pi d^3}\lambda^2 e^{-2\lambda}\left(1-\frac{8\lambda }{3}\alpha+\ldots\right).
\end{equation}
Here for each order in $\alpha$, we only listed the leading term in $\lambda$. Observe that these  terms  are linear in temperature, coming from zero Matsubara frequency, but exponentially small in $\lambda$.

In the large mass low temperature $\vartheta\ll 1\ll \lambda$ region,
\begin{equation}\label{eq6_18_3}
\begin{split}
E_{\text{Cas}}\sim -\frac{3\hbar cA}{16\pi^{\frac{3}{2}} d^3}\lambda^{\frac{3}{2}} e^{-2\lambda}\left(1-\frac{8\lambda }{3}\alpha+\ldots\right),\\
F_{\text{Cas}}\sim -\frac{3\hbar cA}{8\pi^{\frac{3}{2}} d^3}\lambda^{\frac{5}{2}} e^{-2\lambda}\left(1-\frac{8\lambda }{3}\alpha+\ldots\right).
\end{split}
\end{equation}
Similarly,  for each order in $\alpha$, we only listed the leading term in $\lambda$. This terms comes from the zero temperature region. They are exponentially small in $\lambda$. The thermal corrections are exponentially small.

From \eqref{eq6_18_2} and \eqref{eq6_18_3}, we find that in the large mass region, a small deviation from infinite conductivity tends to reduce the strength of the Casimir force.

Next, we turn to the small mass $\lambda\ll 1$ region. In the high temperature $\vartheta\gg 1$ region,
\begin{equation}\label{eq6_18_6}
\begin{split}
E_{\text{Cas}}\sim &-\frac{k_B TA}{8\pi d^2}\zeta_R(3)\left\{ \left[1+\frac{2}{\zeta_R(3)}\lambda^2\ln\lambda +\frac{3}{2\zeta_R(3)}(2\ln 2-1)\lambda^2-\frac{2}{\zeta_R(3)}
\lambda^3+\ldots\right]\right.\\&\left.\hspace{2.5cm}-2\left[1-\frac{1}{\zeta_R(3)}\lambda^2+\frac{4}{3\zeta_R(3)}\lambda^3+\ldots\right]\alpha+\ldots\right\},\\
F_{\text{Cas}}\sim &-\frac{k_B TA}{4\pi d^3}\zeta_R(3)\left\{ \left[1-\frac{1}{ \zeta_R(3)} \lambda^2+\frac{1}{\zeta_R(3)}
\lambda^3+\ldots\right] -3\left[1-\frac{1}{3\zeta_R(3)}\lambda^2 +\ldots\right]\alpha+\ldots\right\}.
\end{split}
\end{equation}Again, these terms are linear in temperature, coming from zero Matsubara frequency. In the massless $\lambda=0$ limit, we find that
\begin{equation}
\begin{split}
E_{\text{Cas}}\sim &-\frac{k_B TA}{8\pi d^2}\zeta_R(3)\left\{  1-2 \alpha+\ldots\right\},\\
F_{\text{Cas}}\sim &-\frac{k_B TA}{4\pi d^3}\zeta_R(3)\left\{1-3 \alpha+\ldots\right\}.
\end{split}
\end{equation}
Setting $\alpha=0$ gives
\begin{equation}
\begin{split}
E_{\text{Cas}}\sim &-\frac{k_B TA}{8\pi d^2}\zeta_R(3),\\
F_{\text{Cas}}\sim &-\frac{k_B TA}{4\pi d^3}\zeta_R(3),
\end{split}
\end{equation}which are well known results for high temperature asymptotics  for perfect conductors \cite{19,9,10}. From \eqref{eq6_18_6}, it is interesting to observe that the existence of small mass and small deviation from infinite conductivity both reduce the strength of the Casimir force. Moreover, the existence of small mass will reduce the effect of small deviation from infinite conductivity.

In the small mass low temperature $\lambda,\vartheta\ll 1$ region, we  consider two cases. When $\vartheta\ll \lambda\ll 1$,
we find that
\begin{equation*}
\begin{split}
E_{\text{Cas}}=&-\frac{\pi^2\hbar cA}{720 d^3}\left\{\left[1-\frac{15}{\pi^2}\lambda^2+\left(\frac{90}{\pi^3}-\frac{80}{\pi^4}\right)\lambda^3+\ldots\right]
-4\left[1-\frac{5}{\pi^2}\lambda^2\right]\alpha+\ldots\right\},\\
F_{\text{Cas}}=&-\frac{\pi^2\hbar cA}{240 d^4}\left\{\left[1-\frac{ 5}{\pi^2}\lambda^2 +\ldots\right]
-\frac{16}{3}\left[1-\frac{5}{2\pi^2}\lambda^2\right]\alpha+\ldots\right\}.
\end{split}
\end{equation*}These terms come from the zero temperature region. The thermal correction terms are exponentially small.

When $\lambda \ll \vartheta \ll 1$, there are polynomial order terms of the thermal correction:
\begin{equation*}
\begin{split}
E_{\text{Cas}}=&-\frac{\pi^2\hbar cA}{720 d^3}\left\{\left[\left(1+\frac{45}{\pi^6}\zeta_R(3)\vartheta^3-\frac{\vartheta^4}{\pi^4}\right) -\frac{15}{\pi^2}\lambda^2\left(1+\frac{6\vartheta}{\pi^2 } \left(\ln\vartheta-\ln\lambda-\ln(2\pi) +\frac{1}{2}\right)- \frac{\vartheta^2}{\pi^2}\right)+\ldots\right]
\right.\\&\hspace{2.5cm}\left.-4\left[\left(1-\frac{45}{2\pi^6}\zeta_R(3)\vartheta^3+\frac{\vartheta^4}{\pi^4}\right)-\frac{5}{\pi^2}\lambda^2
\left(1+\frac{3}{2\pi^4}\zeta_R(3)\vartheta^3\right)+\ldots\right]\alpha+\ldots\right\},\\
F_{\text{Cas}}=&-\frac{\pi^2\hbar cA}{240 d^4}\left\{\left[\left(1+\frac{1}{3}\frac{\vartheta^4}{\pi^4}\right)-\frac{ 5}{\pi^2}\lambda^2\left(1+\frac{\vartheta^2}{\pi^2}\right) +\ldots\right]
\right.\\&\hspace{2.5cm}\left.-\frac{16}{3}\left[\left(1-\frac{45}{8\pi^6}\zeta_R(3)\vartheta^3\right)-\frac{5}{2\pi^2}\lambda^2\left(1-\frac{3}{4\pi^4}\zeta_R(3)\vartheta^3\right)\right]\alpha+\ldots\right\}.
\end{split}
\end{equation*}
In the massless limit $\lambda=0$, we have
\begin{equation}\label{eq6_18_5}
\begin{split}
E_{\text{Cas}}=&-\frac{\pi^2\hbar cA}{720 d^3}\left\{ \left(1+\frac{45}{\pi^6}\zeta_R(3)\vartheta^3-\frac{\vartheta^4}{\pi^4}\right) -4 \left(1-\frac{45}{2\pi^6}\zeta_R(3)\vartheta^3+\frac{\vartheta^4}{\pi^4}\right) \alpha+\ldots\right\},\\
F_{\text{Cas}}=&-\frac{\pi^2\hbar cA}{240 d^4}\left\{ \left(1+\frac{1}{3}\frac{\vartheta^4}{\pi^4}\right) -\frac{16}{3} \left(1-\frac{45}{8\pi^6}\zeta_R(3)\vartheta^3\right) \alpha+\ldots\right\}.
\end{split}
\end{equation}
The latter
 agrees with \cite{3} up to the first order in $\alpha$. Again, we find that the existence of small mass and small deviation from infinite conductivity both reduce the strength of the Casimir force. Moreover, the existence of small mass will reduce the effect of small deviation from infinite conductivity.

\section{Application to Casimir effect in Randall-Sundrum spacetime}
One of the main motivation for studying Casimir effect of massive vector field comes from extra-dimensional physics. To study the Casimir effect of the electromagnetic field in a spacetime with extra dimensions, one can use the Kaluza-Klein decomposition to decompose the electromagnetic field to an infinite tower of massive vector fields in four dimensions. For   two parallel perfectly conducting plates in the Randall-Sundrum spacetime, we have studied the Casimir effect along this line in \cite{11}.

Recall that  the  spacetime underlying the Randall-Sundrum (RS) model is a 5$D$ anti-de Sitter space (AdS$_5$) with background   metric
\begin{equation}\label{eq7_8_1}
ds^2=e^{-2\kappa|y|}\eta_{ab}dx^{a}dx^{b}-dy^2,
\end{equation} where $\eta_{ab}=\text{diag}(1, -1, -1, -1)$ is the usual 4$D$ metric on the Minkowski spacetime $M^{4}$. The extra dimension with coordinate $y$ is  compactified on the orbifold $S^1/\mathbb{Z}_2$. $\kappa$ is a parameter that determines the degree of curvature of the AdS$_5$ space. There are two 3-branes with equal and opposite tensions,   localized at $y=0$ and $y=\pi R$ respectively, where $R$ is the compactification radius of the extra dimension. Using Kaluza-Klein decomposition, an electromagnetic field in the $5D$ Randall-Sundrum spacetime is decomposed into an electromagnetic field (massless vector field) in the $4D$ Minkowski spacetime, and an infinite family of massive Proca fields with masses $m_1< m_2<\ldots$, where
$$z=\frac{m_jc}{\hbar}$$ are positive solutions of the equation \cite{11, 14}:
$$J_0\left(\frac{z}{\kappa} \right)Y_0\left(\frac{z e^{\pi\kappa R}}{\kappa} \right)-Y_0\left(\frac{z}{\kappa} \right)J_0\left(\frac{z e^{\pi\kappa R}}{\kappa} \right)=0.$$
To solve the hierarchy problem between the Planck and electroweak scales, it is required that $\kappa R\simeq 12$ \cite{14}. In this case,
one can show that
$$\frac{m_j c}{\hbar}\simeq \pi \kappa e^{-\pi \kappa R}j.$$
 For the effect of the extra dimension to be significant, we require that
$$\pi \kappa d e^{-12 \pi}\sim 1.$$   For $d\sim 100$nm, this amounts to $\kappa \sim 10^8$ GeV, consistent with the numerical result found in \cite{11} for perfect conductors. For $\kappa> 10^{10}$ GeV,
$$\lambda_j =\frac{m_j c d}{\hbar}\gg 1,$$we find from the asymptotic analysis in the last section that the leading term of the Casimir force due to the Kaluza-Klein nonzero modes is exponentially small. Hence we only see the  $4D$ (massless) electromagnetic Casimir effect. When $\kappa\leq 10^8$ GeV, there will be significant correction to the Casimir force from a finite number of the Kaluza-Klein nonzero modes with
$$\lambda_j =\frac{m_j c d}{\hbar}<1.$$
To get a rough idea of how big the contribution can be, one can  take a finite sum over the asymptotic formulas derived in the previous section.

\section{Conclusion}
In this article, we have derived a generalization of the Lifshitz formula to describe the Casimir interaction between two parallel semi-infinite magnetodielectric slabs separated by a magnetodielectric medium. Specialized to two parallel real metals with dielectric property described by the plasma model, we expand the Casimir energy and the Casimir force up to first order in the finite conductivity correction. Asymptotic behaviors of the Casimir energy and Casimir force in combinations of high/low temperature and large/small mass regions are discussed. It is found that when the mass is large, the Casimir force is exponentially small. In the small mass region, the dominating term is the massless term, and the massive correction   tend to reduce the strength of the Casimir force and the effect of deviation from infinite conductivity. The application to Casimir effect in Randall-Sundrum spacetime is briefly discussed. It is found that for the electromagnetic Casimir effect in the Randall-Sundrum spacetime to be significantly different from the $4D$ electromagnetic Casimir effect, the warping parameter of the Randall-Sundrum spacetime $\kappa$ must satisfy $\kappa \sim 10^8$ GeV, when the separation between the two metals is about 100nm.

A massive vector field has longitudinal modes which satisfy dispersion relation  different from the dispersion relation of the transverse modes. However, the   contribution of the transverse magnetic modes and the longitudes modes cannot be separated. This phenomena has also been observed in the case of two concentric spherical bodies \cite{23}. This makes the Casimir effect of a massive vector field   considerably more complicated than the Casimir effect of an electromagnetic field. In this work, we have only expanded the Casimir energy up to the first order in the finite conductivity correction. However, our scheme can be used to obtain higher order conductivity corrections.

\vspace{1cm}
\noindent
\textbf{Acknowledgement} I would like to thank Professor G. Barton for suggesting me to look into this problem during the QFEXT11 conference. This  work is supported by the Ministry of Higher Education of Malaysia  under the FRGS grant FRGS/2/2010/SG/UNIM/02/2.

\appendix
\section{Asymptotic behaviors of the Casimir energy} \label{A1}

In this section, we compute the asymptotic behaviors of the Casimir free interaction energy in various limits. We consider the TE and TM contribution separately.

\subsection{TE contribution}\label{TE}
Expanding the logarithm in \eqref{eq5_31_3}, we find that
\begin{equation*}
\begin{split}
E_{\text{Cas}}^{\text{TE}}= &-\frac{k_BT A}{2\pi d^2}\sum_{n=0}^{\infty}\!'\sum_{j=1}^{\infty}\frac{1}{j}\int_{\sqrt{z_n^2+\lambda^2}}^{\infty}dz z   \left[R^{\text{TE}}\right]^{2j}e^{-2jz }.\end{split}
\end{equation*}
Since
\begin{equation*}
\left[R^{\text{TE}}\right]^{2j}=1-4\alpha jz +\ldots,
\end{equation*} when $\alpha\ll 1$, we have \begin{equation*}
\begin{split}
E_{\text{Cas}}^{\text{TE}}\sim &\mathcal{A}_0+\mathcal{A}_1\alpha,
\end{split}
\end{equation*}
where
\begin{equation}\label{eq6_11_2}
\begin{split}
\mathcal{A}_0=&-\frac{k_BT A}{2\pi d^2}\sum_{n=0}^{\infty}\!'\sum_{j=1}^{\infty}\frac{1}{j}\int_{\sqrt{z_n^2+\lambda^2}}^{\infty}dz z   e^{-2jz },\\
\mathcal{A}_1=&\frac{2k_BT A}{\pi d^2}\sum_{n=0}^{\infty}\!'\sum_{j=1}^{\infty} \int_{\sqrt{z_n^2+\lambda^2}}^{\infty}dz z^2   e^{-2jz }.
\end{split}
\end{equation}
A straightforward integration gives
\begin{equation}\label{eq6_4_3}
\begin{split}
\mathcal{A}_0=& -\frac{k_BT A}{2\pi d^2}\sum_{n=0}^{\infty}\!'\sum_{j=1}^{\infty}\left(\frac{\sqrt{z_n^2+\lambda^2}}{2j^2}+\frac{1}{4j^3}\right)e^{-2j\sqrt{z_n^2+\lambda^2}},
\end{split}
\end{equation}
\begin{equation}\label{eq6_4_4}
\begin{split}
\mathcal{A}_1=& \frac{k_BT A}{2\pi d^2}\sum_{n=0}^{\infty}\!'\sum_{j=1}^{\infty}\left(\frac{2(z_n^2+\lambda^2)}{j}+\frac{2\sqrt{z_n^2+\lambda^2}}{j^2}+\frac{1}{j^3}\right)e^{-2j\sqrt{z_l^2+\lambda^2}}.
\end{split}
\end{equation}

 Let
$$\vartheta=\frac{2\pi k_B Td}{\hbar c},$$ so that $z_n=n\vartheta$. $\vartheta\gg 1$ corresponds to high temperature limit, and $\vartheta\ll 1$ corresponds to low temperature limit.

In the large mass $\lambda\gg 1$ limits, obviously $\mathcal{A}_0$ and $\mathcal{A}_1$ goes to zero exponentially fast. In addition, if $\vartheta \gg 1$, the leading terms come  from the terms with zero Matsubara frequency. From \eqref{eq6_4_3} and \eqref{eq6_4_4}, we find that in the limit $\lambda, \vartheta\gg 1$,
\begin{equation*}
\begin{split}
\mathcal{A}_0\sim &-\frac{k_BT A}{8\pi d^2}\lambda e^{-2\lambda},\\
\mathcal{A}_1\sim &\frac{k_BT A}{2\pi d^2}\lambda^2 e^{-2\lambda}.
\end{split}
\end{equation*}
When $\vartheta\ll 1\ll \lambda$, the dominating term comes from the zero temperature term, and the thermal correction terms are exponentially small. We have
\begin{equation*}
\begin{split}
\mathcal{A}_0\sim &-\frac{k_BT A}{2\pi d^2} \frac{\hbar c}{2\pi k_B T d}\sum_{j=1}^{\infty}\frac{1}{j}
\int_0^{\infty} du \int_{\sqrt{u^2+\lambda^2}}^{\infty}dz z   e^{-2jz }\\
\sim &-\frac{\hbar c A}{4\pi^2 d^3}\int_{\lambda}^{\infty} dz z\sqrt{z^2-\lambda^2}e^{-2z}\\
=&-\frac{\hbar c A}{4\pi^2 d^3}\int_{0}^{\infty} dz (z+\lambda)\sqrt{z^2+2z\lambda}e^{-2(z+\lambda)}\\
\sim &-\frac{\hbar c A}{4\pi^2 d^3}\sqrt{2}\lambda^{\frac{3}{2}}e^{-2\lambda}\int_0^{\infty}dz\,z^{\frac{1}{2}}e^{-2z}\\
\sim &-\frac{\hbar c A}{16\pi^{\frac{3}{2}} d^3}\lambda^{\frac{3}{2}}e^{-2\lambda}.
\end{split}
\end{equation*}
Similarly, we find that when $\theta\ll 1\ll\lambda$,
\begin{equation*}
\mathcal{A}_1\sim \frac{\hbar c A}{4\pi^{\frac{3}{2}} d^3}\lambda^{\frac{5}{2}}e^{-2\lambda}.
\end{equation*}

Next we consider the small mass $\lambda\ll 1$ limits. In the high temperature  $\vartheta\gg 1$ region, the terms with nonzero Matsubara frequencies go to zero exponentially fast.
Using the formula \begin{equation}\label{eq6_11_1}
e^{-z}=\frac{1}{2\pi i}\int_{c-i\infty}^{c+i\infty} dw\Gamma(w) z^{-w},
\end{equation}
we find that the asymptotic behavior  of $\mathcal{A}_0$ is given by
 \begin{equation*}
\begin{split}
\mathcal{A}_0\sim & -\frac{k_BT A}{4\pi d^2} \frac{1}{2\pi i} \int_{c-i\infty}^{c+i\infty}dw \Gamma(w)
2^{-w}\zeta_R(w+1)\int_{\lambda}^{\infty} dz\,  z^{1-w}\\
 =& -\frac{k_BT A}{4\pi d^2}\frac{1}{2\pi i} \int_{c-i\infty}^{c+i\infty} dw\Gamma(w) 2^{-w}\zeta_R(w+1)\frac{\lambda^{2-w}}{w-2}\\
\sim &-\frac{k_BT A}{16\pi  d^2}\left(\zeta_R(3) + \lambda^2\left(2\ln\lambda+2\ln 2-1\right) -\frac{4}{3}\lambda^3+\ldots\right).
\end{split}
\end{equation*}
The first term is the high temperature leading term of the massless limit.
Similarly, we find that the asymptotic behavior  of $\mathcal{A}_1$ is
\begin{equation*}
\begin{split}
\mathcal{A}_1\sim  \frac{k_BT A}{4\pi  d^2}\left(\zeta_R(3) - \lambda^2  +\frac{2}{3}\lambda^3+\ldots\right).
\end{split}
\end{equation*}

The most interesting case is the low temperature small mass asymptotic behavior. Define
$$\mathcal{Z}(w)=  \sum_{n=0}^{\infty}\!'\frac{1}{(z_n^2+\lambda^2)^{w/2}}.$$
Using \eqref{eq6_11_1}, we obtain from \eqref{eq6_4_3}  that
\begin{equation*}
\begin{split}
\mathcal{A}_0\sim &-\frac{k_BT A}{2\pi d^2} \frac{1}{2\pi i}\int_{c-i\infty}^{c+i\infty} dw\Gamma(w) 2^{-w}\zeta_R(w+1) \frac{\mathcal{Z}\left(w-2\right)}{w-2}
.\end{split}
\end{equation*}
Making a change of variables $w\mapsto w+2$ and using the identity
$$\Gamma(w)=\frac{2^{w-1}}{\sqrt{\pi}}\Gamma\left(\frac{w}{2}\right)\Gamma\left(\frac{w+1}{2}\right),$$ we find that
\begin{equation*}
\begin{split}
\mathcal{A}_0\sim & -\frac{k_BT A}{8\pi^{\frac{3}{2}} d^2}\frac{1}{2\pi i}\int_{c-i\infty}^{c+i\infty}dw  \Gamma\left(\frac{w+3}{2}\right)\zeta_R(w+3)\Gamma\left(\frac{w}{2}\right)\mathcal{Z}(w).
\end{split}
\end{equation*}
Similarly,   we have
\begin{equation*}
\begin{split}
\mathcal{A}_1\sim & \frac{k_BT A}{4\pi^{\frac{3}{2}} d^2}\frac{1}{2\pi i}\int_{c-i\infty}^{c+i\infty}dw (w+2) \Gamma\left(\frac{w+3}{2}\right)\zeta_R(w+3)\Gamma\left(\frac{w}{2}\right)\mathcal{Z}(w).
\end{split}
\end{equation*}

We can consider two cases:   $\vartheta\ll \lambda\ll 1$ and $\lambda\ll \vartheta\ll 1$.
When $\vartheta\ll \lambda $,
\begin{equation*}
\begin{split}
\Gamma\left(\frac{w}{2}\right)\mathcal{Z}(w)=&\sum_{n=0}^{\infty}\!'\int_0^{\infty}dt\, t^{\frac{w}{2}-1}\exp\left(-tn^2\vartheta^2-t\lambda^2\right)\\
=&\frac{\sqrt{\pi}}{\vartheta}\sum_{n=0}^{\infty}\!'\int_0^{\infty}dt\, t^{\frac{w-1}{2}-1}\exp\left(-\frac{\pi^2n^2}{t\vartheta^2}-t\lambda^2\right)\\
=& \frac{\sqrt{\pi}}{\vartheta} \left(\frac{1}{2}\Gamma\left(\frac{w-1}{2}\right)\lambda^{1-w}+2
\sum_{n=1}^{\infty}\left(\frac{\pi n}{\vartheta\lambda}\right)^{\frac{w-1}{2}}K_{\frac{w-1}{2}}\left(\frac{2\pi n\lambda}{\vartheta}\right)\right).
\end{split}
\end{equation*}
Thus, when $\vartheta\ll \lambda\ll 1$,
\begin{equation*}
\begin{split}
\mathcal{A}_0\sim & -\frac{\pi^2\hbar c A}{1440 d^3}\left(1-\frac{15}{\pi^2} \lambda^2+\frac{60}{\pi^3}\lambda^3+\ldots\right),\\
\mathcal{A}_1\sim &  \frac{\pi^2\hbar c A}{240 d^3}\left(1-\frac{5}{\pi^2} \lambda^2+\ldots\right).
\end{split}
\end{equation*}
This comes from the zero temperature term. The thermal correction terms are exponentially small.

When  $\lambda\ll \vartheta $,
\begin{equation*}
\begin{split}
\Gamma\left(\frac{w}{2}\right)\mathcal{Z}(w)=& \frac{1}{2}\Gamma\left(\frac{w}{2}\right)\lambda^{-w}+\sum_{n=1}^{\infty}\int_0^{\infty}dt\, t^{\frac{w}{2}-1}e^{-tn^2\vartheta^2}\sum_{j=0}^{\infty}\frac{(-1)^jt^j}{j!}\lambda^{2j}\\
=&\frac{1}{2}\Gamma\left(\frac{w}{2}\right)\lambda^{-w}+\sum_{j=0}^{\infty} \frac{(-1)^j }{j!}\lambda^{2j}
\Gamma\left(j+\frac{w}{2}\right)\zeta_R(2j+w)\vartheta^{-2j-w}.\end{split}
\end{equation*}
Hence, as $\lambda\ll \vartheta\ll 1$,
\begin{equation*}
\begin{split}
\mathcal{A}_0\sim &-\frac{\pi^2\hbar c A}{1440d^3}\left(\left[1+\frac{45}{\pi^6}\zeta_R(3)\vartheta^3-\frac{\vartheta^4}{\pi^4}\right] -
\frac{15}{\pi^2}\lambda^2\left[1+\frac{6\vartheta}{\pi^2 } \left(\ln\vartheta-\ln\lambda-\ln(2\pi) +\frac{1}{2}\right)- \frac{\vartheta^2}{\pi^2}\right] +\ldots\right),\\
\mathcal{A}_1\sim & \frac{\pi^2\hbar c A}{240d^3}\left(\left[1+ \frac{\vartheta^4}{3\pi^4}\right] -
\frac{5}{\pi^2}\lambda^2\left[1+\frac{\vartheta^2}{\pi^2}\right]   +\ldots\right),
\end{split}
\end{equation*} up to order $\lambda^2$ terms.
\subsection{TM contribution}
For the  TM  contribution, expanding the logarithm in \eqref{eq5_31_5}, we find that
 \begin{equation}\label{eq5_31_6}
\begin{split}
E_{\text{Cas}}^{\text{TM}}=&-\frac{k_BT A}{2\pi d^2}\sum_{n=0}^{\infty}\!'
\sum_{j=1}^{\infty}\frac{1}{j}\int_{\sqrt{z_n^2+\lambda^2}}^{\infty}dz\,z   \left(T_+^j+T_-^j\right)e^{-2jz}.
\end{split}\end{equation}
Up to first order in $\alpha$, we find that
\begin{equation*}
\begin{split}
T_+^j=&1-4\alpha j\frac{zz_n^2}{z^2-\lambda^2}+\ldots,\\
T_-^j= &\left(\frac{\lambda}{z+\sqrt{z^2-\lambda^2}}\right)^{4j}+4j\alpha  z\frac{z_n^2+\lambda^2-z^2}{z^2-\lambda^2}\left(\frac{\lambda}{z+\sqrt{z^2-\lambda^2}}\right)^{4j}+\ldots.
\end{split}
\end{equation*}
Therefore, when $\alpha\ll 1$, we have
\begin{equation*}
\begin{split}
E_{\text{Cas}}^{\text{TM}}\sim &\mathcal{B}_0+\mathcal{C}_0+\left(\mathcal{B}_1+\mathcal{C}_1\right)\alpha,
\end{split}
\end{equation*}
where $\mathcal{B}_0=\mathcal{A}_0$,
\begin{equation}\label{eq6_4_6}
\begin{split}
\mathcal{C}_0=&-\frac{k_BT A}{2\pi d^2}\sum_{n=0}^{\infty}\!'
\sum_{j=1}^{\infty}\frac{1}{j}\int_{\sqrt{z_n^2+\lambda^2}}^{\infty}dz\,z   \left(\frac{\lambda}{z+\sqrt{z^2-\lambda^2}}\right)^{4j}e^{-2jz},\\
\mathcal{B}_1=& \frac{2k_BT A}{ \pi d^2}\sum_{n=0}^{\infty}\!'
\sum_{j=1}^{\infty} \int_{\sqrt{z_n^2+\lambda^2}}^{\infty}dz\,z^2   \frac{z_n^2 }{z^2-\lambda^2}  e^{-2jz},\\
\mathcal{C}_1=&-\frac{2k_BT A}{ \pi d^2}\sum_{n=0}^{\infty}\!'
\sum_{j=1}^{\infty} \int_{\sqrt{z_n^2+\lambda^2}}^{\infty}dz\,z^2   \frac{z_n^2+\lambda^2-z^2}{z^2-\lambda^2} \left(\frac{\lambda}{z+\sqrt{z^2-\lambda^2}}\right)^{4j}e^{-2jz}.
\end{split}
\end{equation}
The asymptotic behaviors of $\mathcal{B}_0$ have been computed in Section \ref{TE}. Here we compute the asymptotic behaviors of $\mathcal{C}_0$, $\mathcal{B}_1$ and $\mathcal{C}_1$.

First we consider the large mass limit $\lambda\gg 1$. In the high temperature $\vartheta\gg 1$ region, the dominating terms are the terms with zero Matsubara frequencies. Making a change of variables $z\mapsto   z+\lambda $, we find that when $\lambda,\vartheta\gg 1$,
\begin{equation}\label{eq6_5_1}
\begin{split}
\mathcal{C}_0\sim &-\frac{k_BT A}{4\pi d^2}
\sum_{j=1}^{\infty}\frac{1}{j}\int_{\lambda }^{\infty}dz\,z   \left(\frac{\lambda}{z+\sqrt{z^2-\lambda^2}}\right)^{4j}e^{-2j z},\\
\sim &-\frac{k_BT A}{4\pi d^2}
 \int_{0 }^{\infty}dz\,(z+\lambda)   \left(\frac{\lambda}{z+\lambda+\sqrt{z^2+2z\lambda}}\right)^{4}e^{-2 (z+\lambda)},\\
 \sim &-\frac{k_BT A}{4\pi d^2}\lambda e^{-2\lambda}
 \int_{0 }^{\infty}dz\,     e^{-2 z}\\
= &-\frac{k_BT A}{8\pi d^2}\lambda e^{-2\lambda}.
\end{split}
\end{equation} Similarly, we have
\begin{equation*}
\begin{split}
\mathcal{B}_1\sim & 0,\\
\mathcal{C}_1\sim &\frac{k_BT A}{2\pi d^2}\lambda^2 e^{-2\lambda}.\end{split}
\end{equation*}

In the region $\vartheta\ll 1\ll \lambda$,
 \begin{equation}\label{eq6_11_4}
\begin{split}
\mathcal{C}_0=&-\frac{\hbar c A}{4\pi^2 d^3}
\sum_{j=1}^{\infty}\frac{1}{j}\int_0^{\infty}du\int_{\sqrt{u^2+\lambda^2}}^{\infty}dz\,z   \left(\frac{\lambda}{z+\sqrt{z^2-\lambda^2}}\right)^{4j}e^{-2jz}\\
=&-\frac{\hbar c A}{4\pi^2 d^3}
\sum_{j=1}^{\infty}\frac{1}{j} \int_{\lambda}^{\infty}dz\,z  \sqrt{z^2-\lambda^2} \left(\frac{\lambda}{z+\sqrt{z^2-\lambda^2}}\right)^{4j}e^{-2jz}\\
\sim &-\frac{\hbar c A}{16\pi^{\frac{3}{2}} d^3}\lambda^{\frac{3}{2}}e^{-2\lambda}.
\end{split}
\end{equation}
Similarly, we find that
\begin{equation*}
\begin{split}
\mathcal{B}_1\sim &\frac{\hbar c A}{12\pi^{\frac{3}{2}} d^3}\lambda^{\frac{5}{2}}e^{-2\lambda},\\
\mathcal{C}_1\sim &\frac{\hbar c A}{6\pi^{\frac{3}{2}} d^3}\lambda^{\frac{5}{2}}e^{-2\lambda}.
\end{split}
\end{equation*}

Next we consider the small mass $\lambda \ll 1$ limits. In the high temperature $\vartheta\gg 1$ region, the dominating term comes from the terms with zero Matsubara frequencies. Therefore it is obvious that $\mathcal{B}_0$ is exponentially small. For $\mathcal{C}_0$,
\begin{equation*}
\begin{split}
\mathcal{C}_0\sim&-\frac{k_BT A}{4\pi d^2}\sum_{j=1}^{\infty}\frac{1}{j}
 \int_{\lambda }^{\infty}dz\,z   \left(\frac{\lambda}{z +\sqrt{z^2-\lambda^2}}\right)^{4j}e^{-2 jz}\\
 \sim &-\frac{k_BT A}{4\pi d^2}\sum_{j=1}^{\infty}\frac{\lambda^2}{j}
 \int_{1 }^{\infty}dz\,z   \left(\frac{1}{z +\sqrt{z^2-1}}\right)^{4j}e^{-2\lambda jz}\\
\sim &-\frac{k_BT A}{4\pi d^2}\frac{1}{2\pi i}\int_{c-i\infty}^{c+i\infty}dw
\Gamma(w)\lambda^{2-w}\sum_{j=1}^{\infty}2^{-w}j^{-w-1}
 \int_{1 }^{\infty}dz\,z^{1-w}   \left(\frac{1}{z +\sqrt{z^2-1}}\right)^{4j}.
\end{split}
\end{equation*}
Making a change of variables
$$z=\frac{1+v}{2\sqrt{ v}},$$ we have
\begin{equation*}
\begin{split}
S(w)=&\sum_{j=1}^{\infty}2^{-w}j^{-w-1}\int_{1 }^{\infty}dz\,z^{1-w}   \left(\frac{1}{z +\sqrt{z^2-1}}\right)^{4j}\\=&
\frac{1}{8}
\sum_{j=1}^{\infty}j^{-w-1} \int_0^{1} dv (1-v)(1+v)^{1-w}  v^{\frac{w}{2}-2+2j}.
\end{split}
\end{equation*}
The largest pole of this function is at $w=-2$.
Thus,
\begin{equation*}
\begin{split}
\mathcal{C}_0\sim & -\frac{k_BT A}{4\pi d^2} \left(\lambda^2 S(0)-\lambda^3S(-1)+ \ldots\right)\\
\sim & -\frac{k_BT A}{4\pi d^2} \left(\lambda^2 \left[\frac{1}{2}\ln 2-\frac{1}{4}\right]-\frac{1}{3}\lambda^3  +\ldots\right)
\end{split}\end{equation*}
Similarly,
\begin{equation*}
\begin{split}
\mathcal{C}_1\sim  &\frac{k_BT A}{ 6\pi d^2}\lambda^3+\ldots.
\end{split}
\end{equation*}

Next we consider the case where $\lambda,\vartheta\ll 1$. When $\vartheta\ll \lambda\ll 1$,
the leading contribution comes from the zero temperature term. As in \eqref{eq6_11_4}, we have
\begin{equation*}
\begin{split}
\mathcal{C}_0\sim & -\frac{\hbar c A}{4\pi^2 d^3}
\sum_{j=1}^{\infty}\frac{\lambda^3}{j} \int_{1}^{\infty}dz\,z  \sqrt{z^2-1} \left(\frac{1}{z+\sqrt{z^2-1}}\right)^{4j}e^{-2\lambda jz}.
\end{split}
\end{equation*}
Similar to the case where $\lambda\ll 1\ll \vartheta$, we find that
\begin{equation*}
\begin{split}
\mathcal{C}_0\sim & -\frac{\hbar c A}{4\pi^2 d^3}
\sum_{j=1}^{\infty}\frac{\lambda^3}{j} \int_{1}^{\infty}dz\,z  \sqrt{z^2-1} \left(\frac{1}{z+\sqrt{z^2-1}}\right)^{4j}+\ldots\\
=&-\frac{\hbar c A}{64\pi^2 d^3}
\sum_{j=1}^{\infty}\frac{\lambda^3}{j}\int_0^1dv (1-v)^2(1+v)v^{2j-\frac{5}{2}}+\ldots\\
=&-\frac{\hbar c A}{24\pi^2 d^3}\left(\pi-\frac{8}{3}\right)\lambda^3+\ldots.
\end{split}
\end{equation*}
For  $\mathcal{C}_1$, one can show in the same way that
 it is $O\left(\lambda^4\ln\lambda\right)$.

 For $\mathcal{B}_1$, we find that as $\vartheta\ll\lambda\ll 1$,
\begin{equation*}
\begin{split}
 \mathcal{B}_1\sim &\frac{\hbar cA}{3 \pi^2 d^3}
\sum_{j=1}^{\infty} \int_{\lambda}^{\infty}dz\,z^2   \sqrt{z^2-\lambda^2} e^{-2jz}\\
=&\frac{\hbar cA}{3 \pi^2 d^3}\lambda^4
\sum_{j=1}^{\infty} \int_{1}^{\infty}dz\,z^2   \sqrt{z^2-1} e^{-2\lambda jz}\\
=&\frac{\hbar cA}{3 \pi^2 d^3} \frac{1}{2\pi i}\int_{c-i\infty}^{c+i\infty}dw\Gamma(w)2^{-w}\lambda^{4-w}\zeta_R(w)
\int_{1}^{\infty}dz\,z^{2-w}   \sqrt{z^2-1} \\
=&\frac{\hbar cA}{12 \pi^{\frac{3}{2}} d^3} \frac{1}{2\pi i}\int_{c-i\infty}^{c+i\infty}dw\Gamma(w)\frac{\Gamma\left(\frac{w-4}{2}\right)}
{\Gamma\left(\frac{w-1}{2}\right)}2^{-w}\lambda^{4-w}\zeta_R(w) \\
=&\frac{\pi^2\hbar cA}{720  d^3}\left(1-\frac{5}{\pi^2}\lambda^2+\ldots\right)
\end{split}
\end{equation*}

The case $\lambda\ll \vartheta\ll 1$ is most technical. After some painstaking computation, one can show that up to order $\lambda^2$, the temperature correction terms are still exponentially small for $\mathcal{C}_0$ and $\mathcal{C}_1$. For $\mathcal{B}_1$, we have
\begin{equation}
\begin{split}
\mathcal{B}_1=& \frac{2k_BT A}{ \pi d^2}\sum_{n=0}^{\infty}\!'
\sum_{j=1}^{\infty} \int_{\sqrt{z_n^2+\lambda^2}}^{\infty}dz\,z^2   \frac{z_n^2 }{z^2-\lambda^2}  e^{-2jz},\\
=& \frac{2k_BT A}{ \pi d^2}\sum_{n=1}^{\infty}
\sum_{j=1}^{\infty} \int_{0}^{\infty}dz\,z \sqrt{z^2+z_n^2+\lambda^2}   \frac{z_n^2 }{z^2+z_n^2}  e^{-2j\sqrt{z^2+z_n^2+\lambda^2}},\\
=&\frac{ k_BT A}{ \pi d^2}\frac{1}{2\pi i}\int_{c-i\infty}^{c+i\infty}dw\Gamma(w)2^{-w}\zeta_R(w)\sum_{n=1}^{\infty} z_n^2
\int_{0}^{\infty}dz  (z+z_n^2+\lambda^2)^{\frac{1-w}{2}}   (z+z_n^2)^{-1}\\
 =&\frac{ k_BT A}{ \pi d^2}\frac{1}{2\pi i}\int_{c-i\infty}^{c+i\infty}dw\Gamma(w)2^{-w}\zeta_R(w)
\\&\hspace{2cm}\times\sum_{n=1}^{\infty} z_n^2 \int_{0}^{\infty}dz  (z+z_n^2)^{-\frac{1+w}{2}}\left(1+\frac{1-w}{2}\lambda^2(z+z_n^2)^{-1} +\ldots\right)   \\
 =&\frac{ k_BT A}{ \pi d^2}\frac{1}{2\pi i}\int_{c-i\infty}^{c+i\infty}dw\Gamma(w)2^{-w}\zeta_R(w)\left(\frac{2}{w-1}\zeta_R(w-3)\vartheta^{3-w}
 -\frac{w-1}{w+1}\lambda^2\zeta_R(w-1)\vartheta^{1-w}+\ldots\right)\\
 =&\frac{\pi^2\hbar c A}{720d^3}\left(\left[1-\frac{90}{\pi^6}\zeta_R(3)\vartheta^3+\frac{3}{\pi^4}\vartheta^4\right]
 -\frac{5\lambda^2}{\pi^2}\left[1-\frac{3}{\pi^2}\vartheta^2+\frac{6}{\pi^4}\zeta_R(3)\vartheta^3\right]+\ldots\right).
\end{split}
\end{equation}

\end{document}